\documentclass[ms]{aastex}
\usepackage{spr-astr-addons}
\usepackage{url}

\RequirePackage{color}

\renewcommand{\vec}[1]{\mbox{\boldmath $#1$}}

\newcommand*\Del{\mathrm{\Delta}}                 

\newcommand{\rmd}{{\ \mathrm d} }

\newcommand{\m}{{\ \mathrm m} }
\newcommand{\s}{{\ \mathrm s} }
\newcommand{\kg}{{\ \mathrm {kg}} }

\newcommand{\J}{{\ \mathrm J} }

\newcommand{\emaila}{wilhelm@mps.mpg.de}
\newcommand{\emailb}{bholadwivedi@gmail.com}

\begin{document}

\title{On the potential energy in a gravitationally bound two-body
system with arbitrary mass distribution}


\shorttitle{}
\shortauthors{K. Wilhelm and B.N. Dwivedi}
\author{Klaus Wilhelm}
\affil{Max-Planck-Institut f\"ur Son\-nen\-sy\-stem\-for\-schung
(MPS), 37077~G\"ottingen, Germany \\ \emaila}
\and
\author{Bhola N. Dwivedi}
\affil{Department of Physics, Indian Institute of Technology
(Banaras Hindu University), Varanasi-221005, India \\
\emailb}


\vspace{1cm}
\begin{abstract}
The potential energy problem in a gravitationally bound two-body system has
recently been studied in the framework of a proposed impact model of
gravitation \citep{WilDwi}. The result was applied to the free fall of the
so-called Mintrop--Ball in G\"ottingen with the implicit assumption that the
mass distribution of the system is extremely unbalanced. An attempt to
generalize the study to arbitrary mass distributions indicated a conflict with
the energy conservation law in a closed system. This necessitated us to
reconsider an earlier assumption made in selecting a specific process out of
two options \citep{Wiletal}. With the result obtained here we can now make an
educated selection and reverse our choice. The consequences are presented and
discussed in detail for several processes. Energy and momentum conservation
could now be demonstrated in all cases.
\end{abstract}
PACS
~~04.20.Cv,    
04.20.Fy,      
04.25.-g,      
04.50.Kd,      
04.90.+e       
%

\section{Introduction}
\label{s:introd}
An earlier study on the potential energy problem \citep{WilDwi} had been
motivated by the remark that the ``potential energy is a rather
mysterious quantity'' \citep{Car98}\footnote{In this context it is of interest
that \citet{Bri65} discussed this problem in relation to electrostatic
potential energy.}. It led to the identification of the
``source region'' of the potential energy for the special case of
a system with two masses~$M_{\rm E}$ and $M_{\rm M}$ subject to the condition
$M_{\rm E} \gg M_{\rm M}$. An attempt to generalize the study without this
condition required either violations of the energy conservation principle
as formulated by \citet{Lau20} for a closed system, or a reconsideration
of an assumption we made concerning the gravitational interaction
process \citep{Wiletal}. The changes necessary to comply with the energy
conservation principle will be the theme of this article, in addition to
a generalization of the potential energy concept for a system of two
spherically symmetric bodies~A
and B with masses~$m_{\rm A}$ and $m_{\rm B}$ without the above condition.

We will again exclude any further energy contributions, such as rotational or
thermal energies, and make use of the fact that the external gravitational
potential of a spherically symmetric body of mass~$m$ and radius~$r_m$ is
that of a corresponding point mass at the centre, i.e.
\begin{equation}
\phi_m(r) = - \, G_{\rm N}\,\frac{m}{r}~~~~~~~~~~ (r > r_m) ~,
\label{eq:potential}
\end{equation}
where $r = |\vec{r}|$ is the distance from the centre of the sphere and
$G_{\rm N} = 6.673\,84(80) \times 10^{-11}\,\m^3\,\kg^{-1}\,\s^{-2}$,
the constant of gravity.\footnote{This and other constants are taken from
2010 CODATA at http://physics.nist.gov/cuu/constants.} The direction of
$\vec{r}$ will be reckoned positive in later calculations.

The energy~$E_m$ and momentum~$\vec{p}$ of a free particle with mass~$m$ moving
with a velocity~$\vec{v}$ relative to an inertial reference system are
related by
%
\begin{equation}
E^2_m - \vec{p}^{\,2}\,c^2_0 = m^2\,c^4_0 ~,
\label{eq:energy}
\end{equation}
where $c_0 = 299\,792\,458\,\m\,\s^{-1}$ (exact)\footnote{Follows from the
definition of the SI base unit ``metre'' (Bureau International des Poids
et Mesures, BIPM, 2006).} is the speed of light in vacuum and
%
\begin{equation}
\vec{p} = \vec{v}\,\frac{E_m}{c^2_0}
\label{eq:momentum}
\end{equation}
\citep{Ein05a,Ein05b}.
For an entity in vacuum with no rest mass ($m = 0$), such as a
photon \citep[cf.][]{Ein05c,Lew26,Oku09}, the energy-momentum
relation in Eq.~(\ref{eq:energy}) reduces to
%
\begin{equation}
E_\nu = p_\nu\,c_0 ~.
\label{eq:photon}
\end{equation}
%

\section{Gravitational impact model and quadrupoles}
\label{s:quadrupoles}
In analogy to Eq.~(\ref{eq:photon}), we have postulated hypothetical
massless entities (named ``quadrupoles'') that obey the energy-momentum
equation
%
\begin{equation}
E_{\rm G} = p_{\rm G}\,c_0 ~,
\label{eq:quadrupole}
\end{equation}
and constructed a gravitational impact model with a background flux
of these entities modified by gravitational centres
\citep[for details see][]{Wiletal}.
Equal absorption and emission number rates
proportional to the mass of a body had been assumed. Nevertheless, the
energy\,--\,and momentum\,--\,change rates must be different in order to
emulate the gravitational attraction in line with Newton's law of gravity. We,
therefore, had to introduce a reduction parameter~$Y \ll 1$ and defined the
energy absorption rate by
%
\begin{equation}
\frac{\rmd E^{\rm ab}_{\rm G}}{\rmd t} =
p_{\rm G}\,c_0\,\frac{\rmd N_m}{\rmd t}
\label{eq:E_ab}
\end{equation}
and the energy emission rate by
%
\begin{equation}
\frac{\rmd E^{\rm em}_{\rm G}}{\rmd t} =
p_{\rm G}\,c_0\,(1 - Y)\,\frac{\rmd N_m}{\rmd t} ~,
\label{eq:E_em}
\end{equation}
where $\rmd N_m/\rmd t = m\,c^2_0/(2\,h)$ is half the intrinsic de Broglie
frequency of a body with mass~$m$.

According to Newton's third law, the interaction rate of quadrupoles with
bodies~A and B must be the same for both bodies even if
$m_{\rm A} \ne m_{\rm B}$:
%
\begin{equation}
\frac{\rmd N_{m_{\rm A},m_{\rm B}}}{\rmd t} =
\frac{\rmd N_{m_{\rm B},m_{\rm A}}}{\rmd t} ~.
\label{eq:ac_re}
\end{equation}
The rate required to emulate Newton's law of gravitation critically depends on
the details of the process.
A spherically symmetric emission of a liberated quadrupole had been assumed
by \citet{Wiletal}. Further studies summarized below have, however,
indicated that an anti-parallel emission with respect to the incoming
quadrupole is more appropriate, because conflicts with the energy and momentum
conversation principles in closed systems can be avoided by the second choice.
It leads to a momentum transfer rate from $m_{\rm A}$ to $m_{\rm B}$ of
$(2 - 3\,Y)\,p_{\rm G}\,\rmd N_{m_{\rm A},m_{\rm B}}/\rmd t$ between
$m_{\rm A}$ and $m_{\rm B}$ by interacting quadrupoles. This implies that
$(2 - Y)\,p_{\rm G}\,\rmd N_{m_{\rm A},m_{\rm B}}/\rmd t$
(that would have been balanced by interactions from the opposite direction)
will not be absorbed by $m_{\rm B}$ from the background.\footnote{For weak
gravitational interactions, the spatial density of quadrupoles with reduced
momentum is very small compared to that of the background.}

Consequently, Eq.~(17) of \citet{Wiletal} has to be modified to a momentum
transfer rate of
%
\begin{eqnarray}
\frac{\rmd P_{m_{\rm A},m_{\rm B}}(r)}{\rmd t} =
- 2\,p_{\rm G}\,Y\,\frac{\rmd N_{m_{\rm A},m_{\rm B}}(r)}{\rmd t} =\nonumber \\
- 2\,p_{\rm G}\,Y\,\kappa_{\rm G}\,\frac{c_0}
{4\,\pi\,h}\,\frac{m_{\rm A}\,m_{\rm B}}{r^2} = 
K_{\rm G}(r) =
- G_{\rm N}\,\frac{m_{\rm A}\,m_{\rm B}}{r^2} ~,
\label{eq:imbalance}
\end{eqnarray}
where $h = 6.626\,068\,57(29) \times 10^{-34}\,\J\,\s$
is the Planck constant and the last two terms are Newton's law of gravitation.

%
\begin{figure}
\begin{center}
\includegraphics[width=\columnwidth]{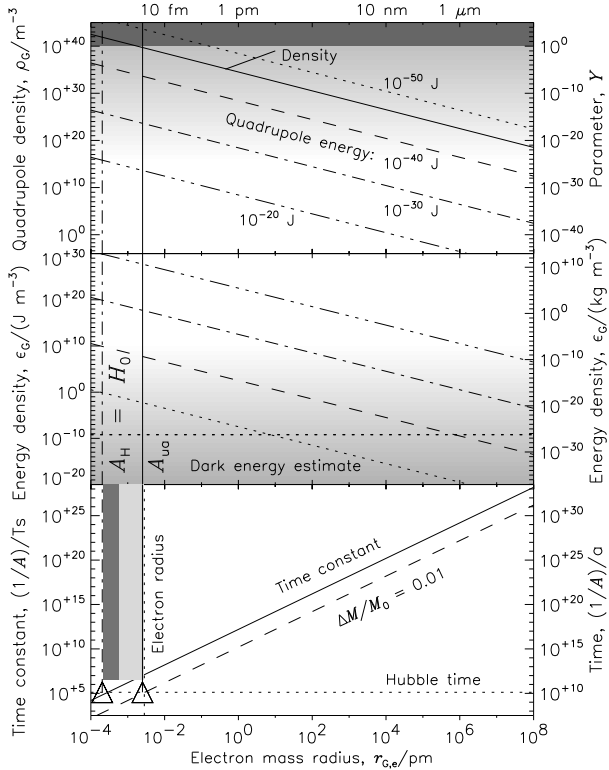}
\end{center}
\caption{The energies of virtual quadrupoles between
$(10^{-50}~{\rm and}~10^{-20})$~J, i.e. below 1\,eV, are assumed as hypothetical
entities of our impact model, as indicated in the upper panel, where
the spatial number and the reduction parameter~$Y$ are plotted as functions
of the electron mass radius~$r_{\rm G,e}$. The range $Y \ge 1$ (dark shading)
is obviously completely excluded by the model, but even values close to unity
are not realistic, because the corresponding energy density in the
lower portion of the middle panel is too small for electron mass
radii~$r_{\rm G,e}$ (triangles in the lower panel, detail are given below).
The cosmic dark energy estimate
\citep{BecMac} is marked in the second panel. It is well below
our acceptable range with quadrupole energies of greater than $10^{-30}$\,J
(unshaded regions in the upper and middle panels). In the
lower panel, the mass accretion time constant and the time required for a
relative mass increase of 1~\% are shown (on the right side in units of years).
Indicated are also the Hubble time, $1/H_0$, as well as the lower limit of the
electron mass radius (left triangle and dark-shaded area) estimated from
the Pioneer anomaly. The light-shaded area
takes smaller Pioneer anomalies into account \citep{WilDwi11}.
It is shown up to the vertical dotted line for the classical electron radius
of 2.82\,fm. The right triangle and the vertical solid line show the
result of a study \citep{WilDwi13a} based on the observed secular
increase of the Sun-Earth distance \citep{KraBru}.
\label{figure_1}}
\end{figure}
The directionally emission assumption reduces the absorption
coefficient by a factor of two, i.e. $\kappa^*_{\rm G} = \kappa_{\rm G}/2$,
and requires, with constant~$\eta_{\rm G}$, a corresponding increase of the
spatial density of quadrupoles to~$\rho^*_{\rm G} = 2\,\rho_{\rm G}$, as well
as twice the quadrupole energy density~$\epsilon_{\rm G}$ in Eq. (20) of
\citet{Wiletal}. All other quantities are not affected, in particular, the
relative mass accretion rate~$A$ will not change.

In Fig.~\ref{figure_1},
we show the new relationships between quadrupole energy, their spatial
density and the corresponding energy density. The extreme logarithmic scale
makes it difficult to see the factor of two, however, we indicate in the
modified figure not only the range of the most likely electron mass
radius~$r_{\rm G,e}$ from our earlier studies \citep{WilDwi11,WilDwi13a}, but
also the parameter ranges of $Y$ and the spatial energy density in the
upper panels that do not yield a realistic impact model
\citep[cf.][]{Wiletal}.

The influence of
a potential shielding effect by a third body placed between two
gravitational centres \citep[cf.][]{Dru97} is also affected and will be
treated in Sect.~\ref{s:perihelion}.

\section{The potential energy}
\label{s:pot_en}

\subsection{Classical mechanics}
\label{ss:Clasic}
We assume that two spherically symmetric bodies~A and B with masses~$m_{\rm A}$
and $m_{\rm B}$, respectively, are placed in space remote from other
gravitational centres at a distance of~$r + \Del r$ reckoned from the position
of~A. Initially both bodies are at rest with respect to an inertial reference
frame represented by the centre of gravity of both bodies. The total energy of
the system then is with Eq.~(\ref{eq:energy}) for the rest energies and
Eq.~(\ref{eq:potential}) for the potential energy
%
\begin{equation}
E_{\rm S} = (m_{\rm A} + m_{\rm B})\,c^2_0 -
G_{\rm N}\,\frac{m_{\rm A}\,m_{\rm B}}{r + \Del r} ~.
\label{eq:total}
\end{equation}
The evolution of the system during the approach of A and B from $r +\Del r$ to
$r$ can be described in classical mechanics.
According to Eq.~(\ref{eq:imbalance}), the attractive force between the
bodies during the approach is approximately constant  for $r \gg \Del r$,
resulting in accelerations of $b_{\rm A} = |K(r)|/m_{\rm A}$ and
$b_{\rm B} = - |K(r)|/m_{\rm B}$, respectively. Since the duration~$t$ of the
free fall of both bodies is the same, the approach of A and B can be
formulated as
%
\begin{eqnarray}
\Del r = s_{\rm A} - s_{\rm B} =
\frac{1}{2}\,(b_{\rm A} - b_{\rm B})\,t^2 = \nonumber \\
\frac{1}{2}\,\left(\frac{1}{m_{\rm A}} +
\frac{1}{m_{\rm B}}\right)\,|K(r)|\,t^2 ~,
\label{eq:acc}
\end{eqnarray}
showing that $s_{\rm A}\,m_{\rm A} = - s_{\rm B}\,m_{\rm B}$, i.e, the centre
of gravity stays at rest. Multiplication of Eq.~(\ref{eq:acc}) by~$|K(r)|$
gives the corresponding kinetic energy equation
%
\begin{eqnarray}
|K(r)|\,\Del r =
\frac{1}{2}\,\left(\frac{K^2(r)\,t^2}{m_{\rm A}} +
\frac{K^2(r)\,t^2}{m_{\rm B}}\right) = \nonumber \\
\frac{1}{2}\,(m_{\rm A}\,v_{\rm A}^2 +
m_{\rm B}\,v_{\rm B}^2) = T_{\rm A} + T_{\rm B} ~,
\label{eq:accel}
\end{eqnarray}
The kinetic energies\footnote{Eqs.~(\ref{eq:energy}) and (\ref{eq:momentum})
together with $E_0 = m\,c^2_0$ \citep{Ein05b} and
$\gamma = 1/\sqrt{1 - v^2/c^2_0}$ yield the relativistic kinetic energy of a
massive body: $T = E - E_0 = E_0\,(\gamma - 1)$. The evaluations for
$T_{\rm A}$ and $T_{\rm B}$ agree in very good approximation with
Eq.~(\ref{eq:accel}) for small $v_{\rm A}$ and
$v_{\rm B}$. \label{footnote}} $T_{\rm A}$ and $T_{\rm B}$
should, of course, be the difference of the potential
energy in Eq.~(\ref{eq:total}) at distances of
$r + \Del r$ and $r$. We find indeed
%
\begin{eqnarray}
- G_{\rm N}\,\frac{m_{\rm A}\,m_{\rm B}}{r + \Del r} +
G_{\rm N}\,\frac{m_{\rm A}\,m_{\rm B}}{r} \approx \nonumber \\
G_{\rm N}\,\frac{m_{\rm A}\,m_{\rm B}}{r^2}\,\Del r = |K(r)|\,\Del r ~.
\label{eq:pot_diff}
\end{eqnarray}
%
\subsection{Quadrupole energy deficiency}
\label{ss:deficiency}
We may now ask the question, whether the impact
model can provide an answer to the ''mysterious'' potential energy problem
in a closed system.
Since the model implies a secular increase of mass of all
bodies, it obviously violates a closed-system assumption. The
increase is, however, only
significant over cosmological time scales, and we can neglect its consequences
in this context. A free single body will, therefore, still be considered as a
closed system with constant mass. In a two-body system both
masses~$m_{\rm A}$ and $m_{\rm B}$ will be constant in such an approximation,
but now there are quadrupoles interacting with both masses.

The number of quadrupoles travelling at any instant of time from one mass to
the other can be calculated from the interaction
rate in Eq.~(\ref{eq:imbalance}) multiplied by the travel time~$r/c_0$
%
\begin{equation}
\Del N_{m_{\rm A},m_{\rm B}}(r) =
\frac{\kappa^*_{\rm G}}{8\,\pi\,h}\,\frac{m_{\rm A}\,m_{\rm B}}{r} ~.
\label{eq:number}
\end{equation}
The same number is moving in the opposite direction.
The energy deficiency of the interacting quadrupoles with respect to the
corresponding background then is together with Eq.~(\ref{eq:E_em}) for each
body
%
\begin{eqnarray}
\Del E_{\rm Q}(r) =
-  p_{\rm G}\,Y\,\kappa^*_{\rm G}\,\frac{c_0}
{8\,\pi\,h}\,\frac{m_{\rm A}\,m_{\rm B}}{r} =
\nonumber \\
- G^*_{\rm G}\,\frac{c_0}{8\,\pi\,h}\,\frac{m_{\rm A}\,m_{\rm B}}{r} =
- \frac{G_{\rm N}}{2}\,\frac{m_{\rm A}\,m_{\rm B}}{r} ~.
\label{eq:deficiency}
\end{eqnarray}
The last term shows\,--\,with reference to Eq.~(\ref{eq:potential})\,--\,that
the energy deficiency~$\Del E_{\rm Q}$ equals {\em half} the potential energy
of body~A at a distance~$r$ from body~B and vice versa.

We now apply Eq.~(\ref{eq:deficiency}) and calculate the difference
of the energy deficiencies for separations of $r + \Del r$ and $r$ for
interacting quadrupoles travelling in both directions and get
%
\begin{eqnarray}
2\,[\Del E_{\rm Q}(r + \Del r) - \Del E_{\rm Q}(r)] = \nonumber \\
- \,G_{\rm N}\,\m_{\rm A}\,m_{\rm B}\,\left(\frac{1}
{r + \Del r} - \frac{1}{r}\right) =
|K(r)|\,\Del r~.
\label{eq:diff_def}
\end{eqnarray}
Consequently, the difference of the potential energies between
$r + \Del r$ and $r$ in Eq.~(\ref{eq:pot_diff}) is balanced
by this difference of the total energy deficiencies.

The physical processes involved can be described as follows:
\begin{enumerate}
\item The number of quadrupoles on their way for a separation of $r + \Del r$
is smaller than that for $r$, because the interaction rate
depends on $r^{-2}$ according to Eq.~(\ref{eq:imbalance}), whereas the
travel time is proportional to $r$.
\item A decrease of~$r + \Del r$ to $r$ during the appraoch of~A and B
increases the number of quadrupoles with reduced energy.
\item The energies liberated by energy reductions are
available as potential energy and are converted into kinetic energies
of the bodies~A and B.
\item With Eqs.~(\ref{eq:energy}) and (\ref{eq:momentum}) and the
approximations in Footnote~\ref{footnote}, it follows that the sum of the
kinetic energies~$T_{\rm A}$ and $T_{\rm B}$, the masses~A and B plus
the total energy deficiencies of the interacting quadrupoles can indeed
be considered to be a closed system as defined by \citet{Lau20}.
\end{enumerate}
Further details of the interaction of quadrupoles with massive particles
have been presented in \citet{Wiletal} explaining the actual conversion
of the liberated energy into kinetic energy.

\section{Multiple quadrupole interactions and the perihelion advances}
\label{s:perihelion}
In large gravitational centres, such as the Sun, multiple
interactions have to be expected before the quadrupoles are emitted with
reduced energy and momentum. The process assumed in \citet{Wiletal} led
to secondary emission centres in the direction of the orbiting body.
Using published data on the secular perihelion advances of the inner planets
of the solar system and the asteroid Icarus, \citet{WilDwi14} found that
the effective gravitational centre is displaced from the centre of the Sun by
approximately 4400~m. Since an analytical derivation of this value from the
mass distribution of the Sun was beyond the scope of the study, the modified
process just has to account, at least in priciple, for such a
displacement.

The proportionality of the linear term in the binomial theorem with the
exponent in
%
\begin{equation}
(1 - Y)^n \approx 1 - n\,Y~~~~~{\rm for}~~Y \ll 1
\label{binom}
\end{equation}
suggests that a linear superposition of the effects of multiple intercations
will be a good approximation, if~$n$ is not too large. Energy reductions
according to Eq.~(\ref{eq:E_em}) are therefore not lost as claimed by
\citet{Dru97}, but they are redistributed to other emission locations within
the Sun. This has two consequences: (1) The total energy reduction is still
dependent on the solar mass, and (2) since emissions from matter closer to the
surface of the Sun in the direction of an orbiting object is more likely to
escape into space than quadrupoles from other locations, the effective
gravitational centre is displaced from the centre of the Sun towards that
object.

\section{Bodies in motion and photon-quadrupole interactions}
\label{s:motion}

\subsection{Moving massive bodies}
\label{ss:massive}
Based on the impact model developed for massive bodies at
rest \citep{Wiletal}, we applied the same concept to bodies in motion
and to photons \citep{WilDwi13b}. The modified interaction process
between quadrupoles and massive bodies presented above leads to the same
consequences in Sect.~2 of that paper, whenever the gravitational absorption
coefficient, the spatial density of the quadrupoles, or the
constant~$p_{\rm G}\,Y\,\kappa$ are concerned. The relations are:
$\kappa^*_{\rm G} = \kappa_{\rm G}/2$,
$\rho^*_{\rm G} = 2\,\rho_{\rm G}$ and
$G^*_{\rm G} = G_{\rm G}/2$. The other quantities and the results are not
affected, because the changes of the gravitational absorption
coefficient and the spatial density of the quadrupoles cancel each other.

\subsection{Photons}
\label{ss:photons}
The deflection of light by gravitational centres according to the
General Theory of Relativity \citep{Ein16} and its observational detection
by \citet{Dysetal} leave no doubt that a photon is deflected by a factor of two
more than expected relative to a corresponding massive particle. Since in
our concept the interaction rate between photons and quadrupoles is twice as
high as for massive particles of the same total energy, the reflection of
a quadrupole from a photon with a momentum of $(1 - Y)\,p_{\rm G}$ must also
be anti-parallel to the incoming one, i.e. a momentum
of~$- 2\,Y\,p_{\rm G}$ will be transferred. Otherwise the correct deflection
angle for photons cannot be obtained.
This modified interaction process has one further important advantage: the
reflected quadrupole can interact with the deflecting gravitational centre
and\,--\,through the process outlined in the paragraph just before
Eq.~(\ref{eq:imbalance})\,--\,transfers~$2\,Y\,p_{\rm G}$, in compliance with
the momentum conservation principle. In the old scheme, the violation of this
principle had no observational consequences, because of the extremely large
masses of relevant gravitational centres, but the adherence to both the
momentum and energy conservation principles is very encouraging and clearly
favours the new concept.

Basically the same arguments are relevant for the longitudinal interaction
between photons and quadru\-poles. The momentum transfer per interaction will
be doubled, but the gravitational absorption coefficient will be reduced
by a factor of two. Together with an increased quadrupole density, all
quantities and results are the same as before. However, a detailed analysis
shows that the momentum conservation principle is now also adhered to.

\section{Conclusions}
\label{s:concl}
In the framework of a recently proposed gravitational impact model
\citep{Wiletal}\,--\,with a modification discussed in this work\,--\,the
physical processes during the conversion of gravitational potential energy
into kinetic energy have been described for two bodies with masses~$m_{\rm A}$
and $m_{\rm b}$ and the source of the potential energy could be identified.
Multiple interactions of quadrupoles leading to shifts of the effective
gravitational centre of a massive body from the ``centre of gravity''
are significantly affected by the modified concept, however, without,
changing the results presented in \citet{WilDwi14}.
The intercation of quadrupoles with photons had to be modified as well, but
the modification did not change the results, with the exception that now
both the energy and momentum conservation principles are fulfilled.


\end{document}